\documentclass[letter]{aa} % for the letters 
\usepackage{graphicx}
\usepackage{natbib}
\usepackage{longtable}
\usepackage[english]{babel}
\usepackage{txfonts}
\usepackage{color}

\begin{document}
   \title{Jet opening angles and gamma-ray brightness of AGN}

%   \subtitle{}

   \author{A.~B.~Pushkarev
          \inst{1,2,3}
          \and
          Y.~Y.~Kovalev\inst{4,1}
	  \and
	  M.~L.~Lister\inst{5}
	  \and
	  T.~Savolainen\inst{1}
%	  \fnmsep
%	  \thanks{{\it Send offprint request to}: A. B. Pushkarev}
          }

   \institute{Max-Planck-Institut f\"ur Radioastronomie, Auf dem H\"ugel 69, 53121 Bonn, Germany\\
              \email{apushkar@mpifr.de}
         \and 
             Pulkovo Astronomical Observatory, Pulkovskoe Chaussee 65/1, 196140 St. Petersburg, Russia
	 \and
             Crimean Astrophysical Observatory, 98688 Nauchny, Crimea, Ukraine
         \and
             Astro Space Center of Lebedev Physical Institute, Profsoyuznaya 86/32, 117997 Moscow, Russia
         \and
             Department of Physics, Purdue University, 525 Northwestern Avenue, West Lafayette, IN 47907, USA
%             \thanks{}
             }

   \date{Received \dots; accepted \dots}

% \abstract{}{}{}{}{} 
% 5 {} token are mandatory
 
  \abstract
  % context heading (optional); leave it empty if necessary  
   {}
   {We have investigated the differences in apparent opening angles between the parsec-scale jets of the active 
   galactic nuclei (AGN) detected by the $Fermi$ Large Area Telescope (LAT) during its first three months of 
   operations and those of non-LAT-detected AGN.}
   {We used 15.4 GHz VLBA observations of sources from the 2 cm VLBA MOJAVE program, a subset of which comprise
    the statistically complete flux density limited MOJAVE sample. We determined the apparent  
    opening angles by analyzing transverse jet profiles from the data in the image plane and by applying a model
    fitting technique to the data in the $(u,v)$ plane. Both methods provided comparable opening angle estimates.}
    {The apparent opening angles of $\gamma$-ray bright blazars are preferentially larger than those of $\gamma$-ray 
    weak sources. At the same time, we have found the two groups to have similar intrinsic opening angle distributions, 
    based on a smaller subset of sources.
    This suggests that the jets in $\gamma$-ray bright AGN are oriented at preferentially smaller angles to the line of
    sight resulting in a stronger relativistic beaming. The intrinsic jet opening angle and bulk flow Lorentz factor 
    are found to be inversely proportional, as predicted by standard models of compact relativistic jets. If a gas 
    dynamical jet acceleration model is assumed, the ratio of the initial pressure of the plasma in the core region 
    $P_0$ to the external pressure $P_\mathrm{ext}$ lies within the range 1.1 to 34.6, with a best fit estimate of 
    $P_0/P_\mathrm{ext}\approx2$.}
   {}


   \maketitle
%
%________________________________________________________________

\section{Introduction}

The EGRET telescope onboard the {\it Compton Gamma Ray Observatory}
\citep{Hartman99} provided detections of $\gamma$-ray emission from many 
extragalactic point sources, most of which were identified with blazars 
\citep[e.g.,][]{M01,SRM03}. The latter term generally refers to objects
classified as flat-spectrum radio-loud quasars and BL Lacs. Although
blazars comprise only a few per cent of the overall AGN population,
they dominate the extragalactic high-energy sky. A systematic
comparison of parsec-scale radio jet structure with $\gamma$-ray
emission in AGN has hitherto been problematic and inconclusive 
due to the limited sensitivity and angular resolution of the EGRET
instrument. For example, a recent analysis of a large 6~cm VLBA Imaging 
and Polarimetry Survey \citep[VIPS;][]{Taylor07} has hinted that 
EGRET-detected blazars might have larger than average jet opening angles, 
but very poor statistics (only four EGRET/VIPS sources) did not allow the 
authors to draw a firm conclusion.

On 2008 June 11, the {\it Fermi Gamma-Ray Space Telescope} (previously
known as {\it GLAST}) was successfully launched by NASA with the Large
Area Telescope (LAT), a successor to EGRET, onboard. The LAT began
operating in August 2008 and has provided $\gamma$-ray observations
with greatly improved sensitivity, superior angular resolution,
large field of view, and wide energy range from about 20~MeV to over 
300~GeV \citep{LAT}. During the first three months of science
operations, the LAT has made 205 high-confidence ($>10\sigma$)
detections of bright $\gamma$-ray sources, 116 of which were associated 
with high galactic latitude ($|b|>10\degr$) AGN (the LAT Bright AGN 
Sample together with 10 lower-confidence associations: LBAS-ext),
as discussed by \cite{LBAS}. Their VLBI radio source identifications 
were reported by \cite{VLBI_LBAS}. The LAT detections of flaring AGN 
have also triggered a number of multiwavelength (from radio to 
$\gamma$-ray) campaigns \citep[e.g.,][]{3C84_Fermi}.

The {\it Fermi} era has already heralded a number of important results
that link the $\gamma$-ray emission and radio properties of
AGN. It has been shown that LAT-detected blazars are brighter and more
luminous in the radio domain at parsec scales \citep{MF2}, have higher
apparent jet speeds \citep{MF1}. The LAT-detected blazars also seem to 
have higher Doppler factors determined from the mm-wavelength variability 
\citep{MF3}. The $\gamma$-ray and radio flares are found to appear within 
a typical timescale of up to a few months and are likely associated with 
the parsec-scale radio core \citep{MF2}. In this Letter we continue 
investigating the relations between $\gamma$-ray brightness and the 
properties of parsec-scale radio jets from the LBAS-ext
list as probed by interferometric observations with the VLBA.
Hereafter we refer to the Fermi LAT 3-month bright source 
list \citep{LBAS} when using the term LAT-detected.

Throughout this letter, we use the term ``core'' as the apparent origin 
of AGN jets that commonly appears as the brightest feature in VLBI images 
of blazars \citep[e.g.,][]{L98,Marscher08}. We use the $\Lambda$CDM 
cosmological model with $H_0=71$~km~s$^{-1}$~Mpc$^{-1}$, $\Omega_m=0.27$, 
and $\Omega_\Lambda=0.73$.

\section{The radio data and source samples}
\label{samples}

\begin{figure*}
\centering
\includegraphics[width=\textwidth,clip=true,angle=0]{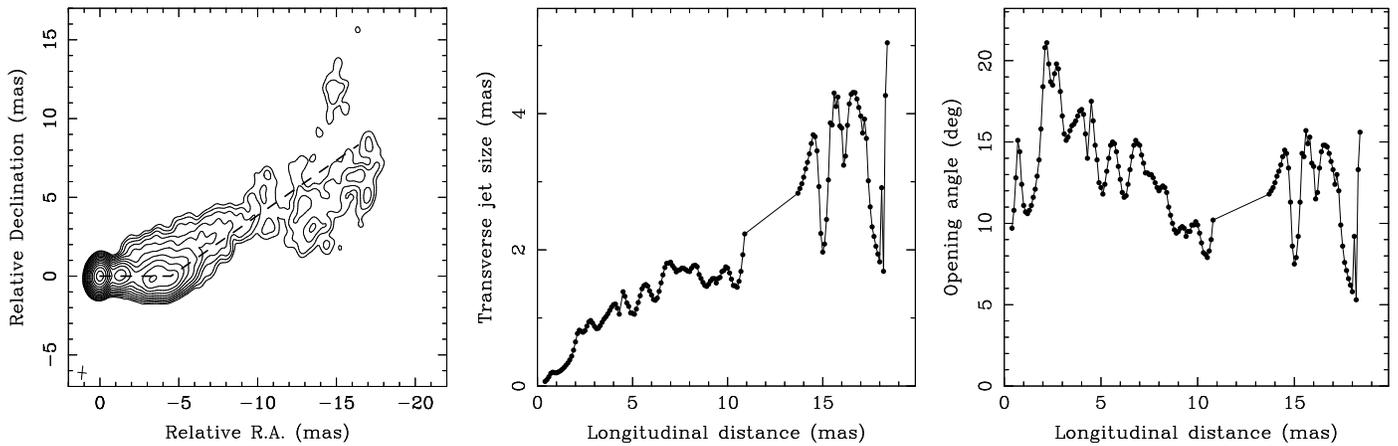}
\caption{
 {\it Left:} Naturally weighted total intensity CLEAN image of 1641+399 
  at 15~GHz observed on 2009 February 25. The jet axis is approximated 
  by two straight lines in position angles of $-90^\circ$ and $-50^\circ$. 
  The contours are plotted at increasing powers of 2, starting from 
  0.4~mJy~beam$^{-1}$. The peak flux density reaches 4770~mJy~beam$^{-1}$. 
  The FWHM of the restoring beam is shown as a cross in the lower left 
  corner. {\it Middle:} Deconvolved FWHM transverse size of the jet along 
  its axis. {\it Right:} Apparent opening angle of the jet along its axis.
 }
 \label{3C345}
\end{figure*}

The MOJAVE program \citep{MOJAVE} is a long-term VLBA program to study 
the structure and evolution of extragalactic relativistic radio jets 
in the northern sky. The full monitoring list currently consists of 
more than 250 sources, and includes a statistically complete, flux-density 
limited sample of 135 AGN (hereafter referred to as MOJAVE-1). All the 
MOJAVE-1 sources have J2000 declination $\delta>-20\degr$ and a 15~GHz 
VLBA correlated flux density $S_\mathrm{c}>1.5$~Jy (2~Jy for $\delta<0\degr$) 
at any epoch between 1994.0 and 2004.0. The weaker radio blazars ($S_c>0.2$~Jy)
detected by {\it Fermi} extend the complete MOJAVE-1 sample to MOJAVE-2.

Although we have analyzed a large portion of the MOJAVE-2 sample, we
focus our statistical analysis on a MOJAVE-1 blazar subset (101
quasars and 22 BL Lacs), which is a dominant fraction of the sample
(91\%). We excluded eight radio galaxies and four optically unidentified 
objects from the MOJAVE-1 sample. Taking into account the fact that the 
LBAS-ext list is restricted in galactic latitude ($|b|>10\degr$), due to the 
decrease of the LAT detection sensitivity at low galactic latitude 
\citep{LBAS}, seven quasars and one BL Lac were excluded from the analysis 
as having $|b|<10\degr$. Of the remaining 115 MOJAVE-1 blazars, 29 are in 
the LBAS-ext list, corresponding to an overall $\gamma$-ray detection rate of
25\%. It is worth noting that the detection rate for BL Lacs is
substantially higher (48\%) compared to that for quasars (20\%) 
for the MOJAVE-1 sample.

\section{Results}
\label{s:results}

The opening angle of a jet in the sky plane can be estimated either
by analyzing transverse jet profiles in the image plane, or by model 
fitting the interferometric visibility data in the {\it(u,v)} plane. 
To increase the robustness of our conclusions we used both methods 
and compared the results.

\subsection{Opening angles from the image plane}
\label{jetcuts}

We used the 15~GHz naturally weighted
MOJAVE\footnote{http://www.physics.purdue.edu/MOJAVE} VLBA
images from the most recent epoch available. This covers the period 
2008--2009 for all but two sources. The opening angle of the jet was
calculated as the median value of $\alpha=2\arctan[0.5(d^2-b_\varphi^2)^{1/2}/r]$,
where $d$ is the full width at half maximum (FWHM) of a Gaussian fitted 
to the transverse jet brightness profile, $r$ is the distance to the core 
along the jet axis, $b_\varphi$ is the beam size along the position angle 
$\varphi$ of the jet-cut, and the quantity $(d^2-b_\varphi^2)^{1/2}$ is 
the deconvolved FWHM transverse size of the jet. The direction of the 
jet axis was determined using the median position angles of all jet
components over all the epochs from model fitting (see next section). 
The slices were taken at 0.1~mas intervals starting from the
position of the VLBI core and continuing up to the region in which the
jet either substantially curved or became undetectable. The ridge
lines for 15 MOJAVE-1 and 6 MOJAVE-2 sources with notably bending jets
were approximated by two straight lines. Opening angle values were
calculated using only those slices that had a peak of the fitted
Gaussian larger than 4 times the rms noise level of the image. In
Fig.~\ref{3C345} the 15 GHz total intensity map of 3C~345 (1641+399),
together with the deconvolved size and opening angle of the jet as a
function of angular distance to the core is shown as an example.

\begin{figure}
 \resizebox{\hsize}{!}{\includegraphics[height=0.5\textwidth,angle=-90,clip=true]{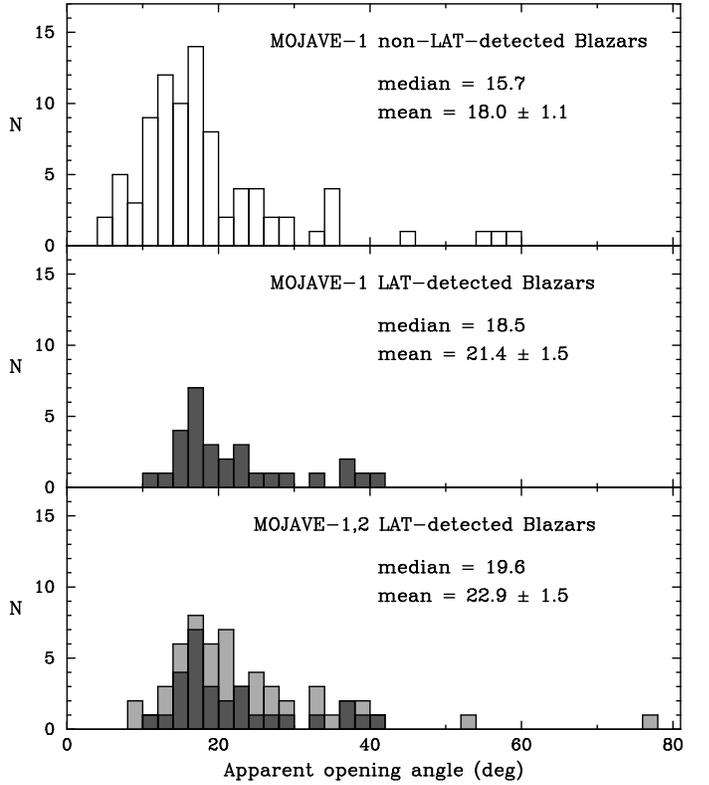}}
 \caption{ Distributions of the apparent opening angle from jet-cut
 analysis for 86 non-LAT-detected ({\it top panel, unshaded}), 29
 LAT-detected ({\it middle and bottom panel, dark gray}) MOJAVE-1
 blazars, and 27 LAT-detected ({\it bottom panel, light gray}) additional
 MOJAVE-2 blazars.} \label{jetcut}
\end{figure}

The distributions of measured opening angle are shown in
Fig.~\ref{jetcut}. The distribution for LAT-detected sources is
narrower and missing the small opening angle sources if compared to
non-LAT-detected sample. A Kolmogorov-Smirnov (K-S) test indicates a
probability of only $p=0.019$ for these two samples being drawn from
the same parent population. If we add 27 additional
LAT-detected blazars from the extended MOJAVE-2 sample the confidence level increases 
to 99.9\% ($p<0.001$) and the mean value to $22\fdg9\pm1\fdg5$ 
(Fig.~\ref{jetcut}, {\it bottom panel}). The jet-slice opening angles 
are on average larger for the 29 LAT-detected MOJAVE-1 blazars in 
comparison with the 86 non-LAT-detected ones, with a mean value of 
$21\fdg4\pm1\fdg5$ versus $18\fdg0\pm1\fdg1$, respectively. 
A Student's T-test confirms that the means are different at 96.1\% 
confidence level for the complete MOJAVE-1 sample and at 99.4\% 
confidence level for the MOJAVE-1,2 sample.
The relatively small difference in mean values is due to a significant 
number of $\gamma$-ray weak sources with very large apparent opening 
angles. In fact, this is compatible with the findings of \cite{MF3}, 
who show that there should be a number of $\gamma$-ray weak sources 
that are viewed nearly end-on and which therefore have very large 
apparent opening angles. Our results support this. We note that the 
TANAMI group, on the basis of VLBI analysis of a Southern hemisphere 
sample of {\it Fermi} detected AGN jets, has also found indications 
of preferentially wider apparent opening angles in 
LAT-detected sources (R.~Ojha et al., in preparation). The calculated values
of the apparent opening angles are listed in 
Table~1\footnote{Table~1 is available in electronic form at
http://www.aanda.org}.

We have studied the effects of a possible bias by BL~Lac objects, 
as discussed by \cite{LBAS}, on our results by comparing the 
apparent opening angle distributions of 22 BL Lacs and 101 
quasars from the complete MOJAVE-1 sample. The K-S test 
indicated no significant difference ($p=0.740$).  
Another check was performed by excluding the BL~Lacs from our 
analysis. In this case, the small set of LAT-detected MOJAVE-1 
quasars ($N = 19$) did not produce a statistically significant 
result. However, comparison of the apparent opening angle distribution 
for LAT-detected versus non-LAT detected quasars in the larger 
MOJAVE-1+MOJAVE-2 sample led to the following. The distributions are 
significantly different according to both the K-S and the Student's 
T-tests ($p_\mathrm{K-S}=0.006$, $p_\mathrm{T-test}=0.017$). We conclude 
therefore that the presence of BL~Lacs does not bias our analysis.

At the same time, we found no significant correlation between the
apparent jet opening angle and average {\it Fermi} LAT 100~MeV~-- 1
GeV photon flux \citep{LBAS}, since the latter depends on the distance
to a source. However, we would expect to see a positive correlation between
the apparent opening angle and $\gamma$-ray luminosity, when the
$\gamma$-ray energy fluxes become available in the next {\it Fermi}
data release.

\subsection{Opening angles from the $(u,\upsilon)$ plane}

The estimates of the opening angle can also be derived from the data
in the $(u,\upsilon)$ plane using a model fitting approach.  The
observed brightness distribution of each source was modeled by a
limited number of two-dimensional Gaussian components using the
``modelfit'' task in Difmap \citep{difmap}. The parameters of the
models are tabulated by \cite{MVI}. We used all available 2336 model
components for 115 MOJAVE-1 blazars with galactic latitude
$|b|>10\degr$. After measuring the opening angle for each jet
component, we averaged them to get one value for each epoch, and then
averaged over the epochs to obtain a final estimate. The statistical
results of the model-fit opening angle distribution analysis are in
good agreement with those from the transverse jet profile method,
confirming that our conclusions are robust. The median value of the
ratio of apparent opening angle derived from jet-cut analysis to that
from the model fitting method is 0.9.
The apparent opening angle distributions of LAT-detected and 
non-LAT-detected MOJAVE-1 blazars are different at 99\% confidence 
level, according to a K-S test.

\subsection{Intrinsic opening angles}

The on-average wider apparent opening angles in LAT-detected
sources can be a consequence of (i) their smaller viewing angles (due
to a projection effect) and/or (ii) their intrinsically different
properties, for example, the presence of spine-sheath structures like
those observed in the $\gamma$-ray bright quasar 1055+018 \citep{ARW99}, 
TeV BL~Lac object Mrk501 \citep{Pushkarev_etal05, Giroletti08} and radio 
galaxy M87 \citep{Kovalev_M87}.

To distinguish between these two possibilities, we have derived the
values of the viewing angle $\theta$ and the bulk Lorentz factor
$\Gamma$ for MOJAVE-1 blazars (see Table~1) using the following relations: 
$$
\theta = \arctan\frac{2\beta_\mathrm{app}}{\beta^2_\mathrm{app}+\delta_\mathrm{var}^2-1}\,,
\quad
\Gamma = \frac{\beta_\mathrm{app}^2+\delta_\mathrm{var}^2+1}{2\delta_\mathrm{var}}\,,
$$ where $\beta_\mathrm{app}$ is the fastest measured radial,
non-accelerating apparent jet speed from the MOJAVE kinematic analysis
\citep{MVI} and $\delta_\mathrm{var}$ is the variability Doppler factor
from the Mets\"ahovi AGN monitoring program \citep{Hovatta09}. 
The overlap of the MOJAVE and Mets\"ahovi programs comprises 56 blazars 
with measured speeds and Doppler factors. 
Indeed, the viewing angles of 21 LAT-detected sources turned out to be 
slightly smaller, with mean value of $3\fdg6\pm0\fdg4$ vs. $5\fdg7\pm1\fdg3$ 
for the non-LAT-detected sources. However, the difference is not statistically 
significant for this small sample (56 sources compared to 115 in the 
apparent opening angle analysis) containing a mixture of quasars and 
BL~Lacs. If only quasars are considered, a marginally significant 
difference ($p = 0.06$) in viewing angle distributions is found between 
LAT-detected and non-LAT-detected sources \citep{MF3}. 
In addition, we have found an indication for BL~Lacs to have on-average 
wider intrinsic opening angles ($2\fdg4\pm0\fdg6$) than those of quasars 
($1\fdg2\pm0\fdg1$). The corresponding distributions are different at confidence 
levels of 94.6\% according to the K-S test, the average values differ with a 
96.1\% confidence according to the Student's T-test.

\begin{figure}
\begin{center}
 \resizebox{0.97\hsize}{!}{\includegraphics[angle=-90,clip=true]{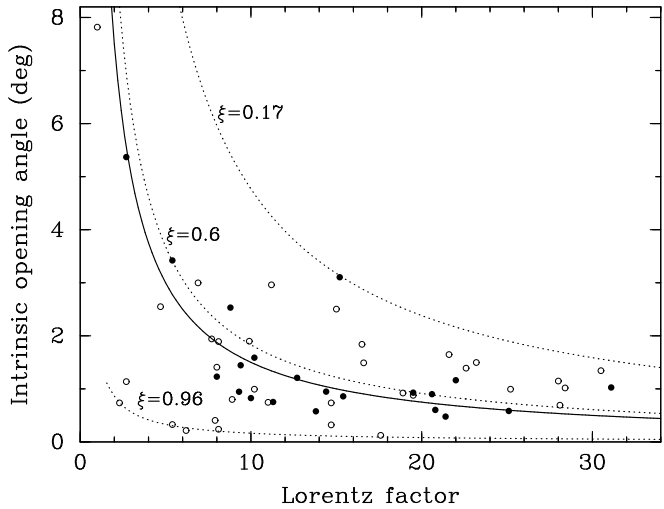}}
\end{center}
 \caption{
          Intrinsic opening angle vs. Lorentz factor for 56 jets. Two sources with  
	  $\Gamma$,\,$\alpha_\mathrm{int}$ values of (45,\,0.96) and (65,\,0.53) 
	  respectively are beyond the plot limits. The solid line shows the median 
	  curve fit with the assumed relation $\alpha=\rho/\Gamma$, where 
	  $\rho$ is a constant. Filled circles correspond to LAT-detected sources, 
	  while open ones correspond to non-LAT-detected sources. The dotted curves 
	  represent relationships between the opening angle and Lorentz factor as 
	  predicted by the gas dynamical model for different values of the parameter 
	  $\xi$.
         }
 \label{oa_vs_gamma}
\end{figure}

To test the second scenario, the intrinsic opening angles
$\alpha_\mathrm{int} = \alpha_\mathrm{app}\,\sin\theta$ were
calculated for the 56 sources (Table~1). The estimated values of 
this parameter range up to $8\degr$ (Fig.~\ref{oa_vs_gamma}). A K-S test 
indicated no significant difference ($p=0.797$) between the samples of
LAT-detected and non-LAT-detected sources, suggesting that the established 
systematic difference in apparent opening angles is most probably the result 
of projection effects, i.e., the $\gamma$-ray bright jets are aligned
closer to our line of sight. 

We have also analyzed the observed dependence between the intrinsic
opening angle and Lorentz factor (Fig.~\ref{oa_vs_gamma}), which are
expected to be inversely proportional according to simple
hydrodynamical models of relativistic jets
\citep{BlandfordKonigl79}. Both the gas dynamical model
\citep{Daly88} and magnetic acceleration models \citep{Komissarov07}
also predict this relation. The observed dependence was fitted
assuming a relation $\alpha_\mathrm{int}=\rho/\Gamma$ with the
coefficient $\rho$ left as a free parameter. The best fit value of
$\rho$ was found to be 0.26~rad by fitting the median curve.
(Fig.~\ref{oa_vs_gamma}, {\it solid line}). Even though the flow
Lorentz factor and the viewing angle (which is needed to determine
$\alpha_\mathrm{int}$), are both calculated from the same observables
($\beta_\mathrm{app}$ and $\delta$), a Monte Carlo simulation shows that
the apparent negative correlation between $\alpha_\mathrm{int}$ and
$\Gamma$ is not due to this degeneracy, but is a genuine effect. 

In the gas dynamical model of compact relativistic jets suggested by
\cite{Daly88}, the opening angle of a jet is a function of the Lorentz
factor and a ratio of the external pressure $P_\mathrm{ext}$ to the
initial pressure $P_0$ of the plasma in the core region,
$\xi=\sqrt{P_\mathrm{ext}/P_0}$. We applied this model for different
values of $\xi$ (Fig.~\ref{oa_vs_gamma}, {\it dotted curves}) and
were able to constrain the parameter $\xi$ to lie within a range
[0.17,\,0.96] with the best fit estimate $\xi=0.67$, corresponding to 
a range of [1.1,\,34.6] for $P_0/P_\mathrm{ext}$ with the best fit estimate 
$P_0/P_\mathrm{ext}\approx2$. Our results confirm those obtained earlier 
by \cite{J05} using a smaller sample of 15 blazar jets.

\section{Summary}
\label{s:summary}

We have measured the projected jet opening angles on parsec scales for 
115 blazars (29 LAT-detected and 86 non-LAT-detected) from the complete 
flux-density limited MOJAVE-1 sample and for 27 additional LAT-detected 
sources monitored by the MOJAVE program. The apparent opening angles 
for $\gamma$-ray bright sources are on average larger than those in
$\gamma$-ray weak ones, while the intrinsic opening angle distributions 
based on smaller samples are statistically indistinguishable. We interpret 
this as an evidence for $\gamma$-ray bright blazars to have preferentially 
smaller viewing angles and, consequently, stronger relativistic beaming, 
which boosts emission in both, the $\gamma$-ray and radio bands. This 
conclusion is consistent with recently obtained results that show 
LAT-detected blazars to be brighter and more luminous in radio domain 
\citep{MF2}, to have faster jets \citep{MF1} and higher variability Doppler 
factors \citep{MF3}.
There is an indication for BL~Lac objects to have on-average wider 
intrinsic opening angles than those of quasars.

The intrinsic opening angle and Lorentz factor are found to be inversely
proportional for a sample of 56 AGN jets in agreement with theoretical
predictions of simple gas dynamical and magnetic acceleration models. The
best approximation of the inferred relation is achieved assuming a ratio
of internal and external jet pressure $P_0/P_\mathrm{ext}\approx2$
in gas dynamical models of compact relativistic jets.

\begin{acknowledgements}

We would like to thank E.~Ros, A.~P.~Lobanov, K.~I.~Kellermann, D.~C.~Homan, 
M.~H.~Aller, M.~H.~Cohen, M.~Kadler, and the rest of the MOJAVE team for the 
usefull discussions. We thank the anonymous referee for useful comments which 
helped to improve the manuscript. This research has made use of data from the 
MOJAVE database that is maintained by the MOJAVE team \citep{MOJAVE}. The 
MOJAVE project is supported under National Science Foundation grant AST-0807860 
and NASA {\it Fermi} grant NNX08AV67G. T.~S. is a research fellow of the Alexander 
von Humboldt Foundation. T.~S. also acknowledges a support by the Academy 
of Finland grant 120516. Y.~Y.~K. was partly supported by the Russian 
Foundation for Basic Research (project 08-02-00545). The VLBA is a facility 
of the National Science Foundation operated by the National Radio Astronomy 
Observatory under cooperative agreement with Associated Universities, Inc.

\end{acknowledgements}

\bibliographystyle{aa}
\bibliography{pushkarev}

\end{document}